# Query Expansion for Patent Searching using Word Embedding and Professional Crowdsourcing

**Arthi Krishna, Ye Jin, Christine Foster, Greg Gabel, Britt Hanley and Abdou Youssef**

United States Patent and Trademark Office, Alexandria, VA, USA
{Arthi.Krishna, Ye.Jin, Christine.Foster, Greg.Gabel, Britt.Hanley}@uspto.gov and ayoussef@gwu.edu

**Abstract**

The patent examination process includes a search of previous work (referred to as "prior art") to verify that a patent application describes a novel invention. Patent examiners primarily use keyword-based searches to uncover prior art. A critical part of keyword searching is query expansion, which is the process of including alternate terms such as synonyms and other related words, since the same concepts are often described differently in the literature. Patent terminology is often domain specific. By curating technology-specific corpora and training word embedding models based on these corpora, we are able to automatically identify the most relevant expansions of a given word or phrase. We compare the performance of several automated query expansion techniques against expert specified expansions. Furthermore, we explore a novel mechanism to extract related terms not just based on one input term but several terms in conjunction by computing their centroid and identifying the nearest neighbors to this centroid. Highly skilled patent examiners are often the best and most reliable source of identifying related terms. By designing a user interface that allows examiners to interact with the word embedding suggestions, we are able to use these interactions to power crowdsourced modes of related terms. Learning from users allows us to overcome several challenges such as identifying words that are bleeding edge and have not been published in the corpus yet. This paper studies the effectiveness of word embedding and crowdsourced models across 11 disparate technical areas.

## Introduction

Applications for patents submitted to the United States Patent and Trademark Office (USPTO) are reviewed by patent examiners in order to ensure the validity of approved patents. The patent examination process includes a search of previous work, referred to as "prior art," in order to verify that the application describes a novel invention. Patent examiners typically use keyword searches to retrieve relevant prior art. Query expansion, including through the use of synonyms and other related terminology, is often critical to successful recall of relevant prior art, since the same concepts are often described using disparate terminology in the literature. In addition, patent searchers often construct search queries that include both broader and narrower terms. As an example, a patent application claiming a broader genus (or hypernym term) such as "salt" cannot be allowed where there is prior art disclosure of a narrower species (or hyponym term) such as "sodium chloride". Concomitantly, patent searchers often employ broader, hypernym terms in their search queries to ensure effective recall of all relevant prior art. As patentable inventions must be not only novel but also non-obvious, there is a recognized need for effective query expansion that does not confine the patent search to the exact terms or phrases that appear in a patent application.

As patent examiners gain experience, they become experts at identifying the underlying concepts of a patent application and predicting the words and variations of words that would retrieve relevant prior art. Some examiners may maintain informal compilations of commonly searched words along with their corresponding alternate representations, but such efforts are sporadic and not readily amenable to knowledge sharing. Technology-specific resources exist for some domains (such as Medical Subject Headings (MeSH) or IEEE thesaurus), but these require user awareness and reliance on multiple different websites/resources, and may not be consistently curated or updated. Large-scale manual curation of a "patent thesaurus" would be prohibitively labor-intensive, given the many different technological fields, as well as the continuously evolving technology found in patent applications.

In the current study, we present a three-part approach for assisting in search query expansion in patent searching: 1. Exploring alternate machine learning-based identification of related terms; 2. Training models based on a generic patent corpus vs. corpora of patent documents clustered into discrete technological fields; and 3. Crowdsourcing by professional patent examiners. We will also summarize the lessons learned in deploying AI solutions to government and follow up with a specific use case of how we have taken advantage of our insight - retrieving related terms based on several user-selected terms by computing the nearest neighbors of the centroid of the terms in vector space.

## Related Work

Automation of query expansion in the patent sphere has been previously explored, although to our knowledge no existing approach is robust enough to replace the need for manual search query generation by an individual patent searcher. Nanba extracted synonyms from the "Description of Symbols" field for patent documents that were related by citations, and thus likely to be within the same technological field (Nanba 2007). Automatic generation of candidate synonym sets showed potential promise as an "on-demand" resource for suggesting possible related terms (e.g., to a patent examiner constructing a search query string), while other approaches using Pseudo Relevance Feedback or the WordNet® thesaurus were not sufficiently effective means of query expansion in the context of patent searching (Ganguly et. al. 2011). Tannebaum, Mahdabi and Rauber (2015) explored semi-automatic query term expansion for patent search using USPTO examiner search query logs.

Our approach is to improve query expansion not just as a means to improve patent search but as an end goal in itself. At present, there are not many datasets available to validate the quality of patent synonyms. As a part of our contribution, we have collected patent examiner-curated synonym datasets, named PatSynSet. For each field studied, we have collected a list of the top 20 terms that commonly require augmentation with alternate words and phrases. Initially, examiners were instructed to provide a comprehensive set of alternates for each of the 20 terms. However, our analysis revealed that these initial lists were not exhaustive and that commonly, some additional terms suggested to the examiners were also judged by the examiners to be useful search equivalents. In order to improve the quality of the dataset, examiners were asked to manually review additional terms suggested by thesauri in their fields, synonyms recovered from search logs, and terms suggested by colleagues. Though the PatSynSet data is not yet large enough to stand on its own as a training set, it is effective as a test set with cross-domain representation. We are in the process of expanding the dataset as a part of our ongoing efforts.

## Word Embedding to Identify Potential Related Terms

Word embedding techniques, where words are represented as vectors that capture semantics relations between each other, allow us to identify terms related to a given word or phrase. We trained two alternate word embedding models using corpora of patent documents in the technological field of immunology (Workgroup 1640). The first method was to train skip-gram model of word2vec (Mikolov et. al. 2013) made available through TensorFlow version 1.12.0 (https://www.tensorflow.org/tutorials/representation/word2vec).

The second used fastText (Bojanowski et. al. 2017) version 0.2.0 with the following hyperparameters: minCount – 5, wordNgrams – 3, -bucket-2000000, minn -1. Patent data from 2001 to 2018 was used as the training corpus. Training for both approaches were run on m5.12xlarge AWS instances (48 vCPU and 192 Memory (GiB).

The synonym sets from PatSynSet were used as the testing dataset, and the values of related terms from both word2vec and fastText models were assessed using precision, recall and F1 scores. Though the results of TensorFlow word2vec were comparable to those of fastText, the ability to train fastText at almost twice the speed led to the choice of fastText due to resource constraints. In this effort, we did not create any other custom machine learning software, making use only of available open source systems.

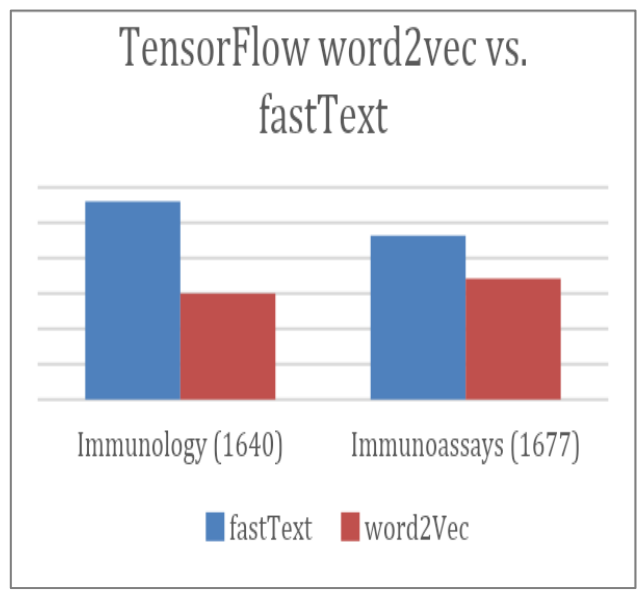

*Figure 1 shows comparison of the F1 scores of word2vec and fastText models trained on the same corpus evaluated against PatSynSet*

We also compiled a corpus for a smaller subset within immunology, corresponding to a single USPTO Art Unit (Art Unit 1677), which examines patent applications in the area of specific binding assay methods and devices (e.g., immunoassays and antibody-based test strips).

**Pre-processing Corpus Data**

The patent data was pre-processed using tokenization with the Natural Language Toolkit. In this process, non-ASCII characters were removed, patent-specific stop words were scrubbed, and punctuation marks were replaced by white space. Several additional domain-specific refinements were also required. The immunology corpora, for instance, required custom code to suppress retrieval of biological sequences (amino acid sequences and nucleic acid sequences), as it was found that such sequences were often spuriously identified as related terms. The pre-processing of chemical formula also required custom code.

**Training Word Embedding with Technology-specific Corpora**

Incoming applications for patents filed with the USPTO are classified by subject matter according to the Cooperative Patent Classification (CPC) system and routed to one of several Technology Centers that best matches the subject matter of the invention. For example, Technology Center 1600 examines applications involving Biotechnology and Organic Chemistry, while Technology Center 2800 examines applications involving Semiconductors/Memory, Optics/Photocopying, Electrical Circuits & Systems and Printing/Measuring & Testing. Within each Technology Center, there are several "Workgroups" made up in turn of individual "Art Units" that are staffed by patent examiners with expertise in a specific field. It is noted that individual Art Units may have clear correspondence with certain areas of the CPC, although this is not necessarily the case due to the multidisciplinary and continually evolving nature of technology.

We hypothesize that by using a set of documents most related to the user's technology (say a particular CPC code or Art Unit) to train word embedding models, we will be able to generate a model that most accurately captures related terms of that field. In the pilot, we had 11 examiners from non-overlapping fields provide 20 most frequent concepts they searched along with a list of related words and phrases. Fig 2. shows the F1 scores of how well a customized word embedding model performed against PatSynSet, as compared with the score of a generic patent model. As shown, accuracy improves when using a technology-specific patent corpus.

The use of technology-specific corpora was able to address word sense disambiguation, which can be a significant challenge in patent searching. As one example, "mold" in a patent application pertaining to microbiology would likely have an entirely different meaning than "mold" in an application pertaining to 3D printing. We also found that mining potential related terms directly from a technology-specific patent corpus using word representation models was effective in retrieving not only synonyms but also more distantly related hypernym and hyponym terms, which are valuable in the highly recall-oriented task of patent searching. This approach is also fast, automated, and may be readily scaled to additional technological fields. Nevertheless, the quality of the raw model results could not approach those generated manually by professional patent searchers.

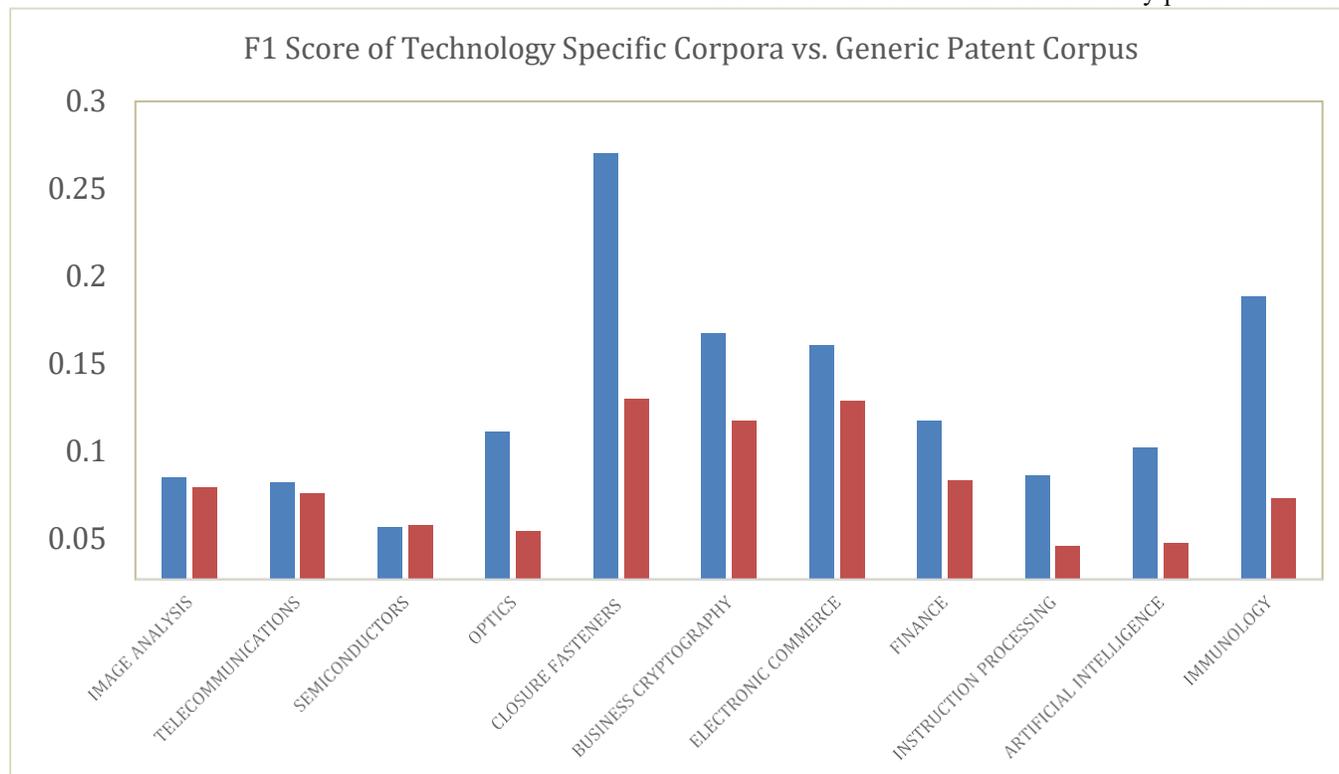

*Figure 2 Related terms retrieved from fastText models trained on technology specific corpus is compared against PatSynSet as is a model trained on an undifferentiated patent corpus*

### Crowdsourcing by Patent Professionals

Crowdsourcing has been previously applied to the task of thesaurus generation. Starting from existing dictionaries, Braslavski, Ustalov and Mukhin (2014) outline a process for Russian thesaurus creation in which noun synsets are edited by native Russian speakers. Krek, Laskowski and Robnik-Šikonja (2017) performed word co-occurrence network analysis on an English-Slovenian dictionary as a starting point for generation of a Slovene thesaurus, which they then coupled with crowdsourcing techniques.

As noted in the previous section, even with the improvements of using domain specific corpus, the accuracy patent examiner experts. By developing mechanisms for capturing user feedback (such as user interface that can capture user interactions that refine synonyms), we are able to leverage this data to generate better crowdsourced results. Fig 3. shows the user interface of a custom program designed for this purpose of collecting data. As shown, each user (name shown in top-right corner) is associated with a profile that presents a list of available technology specific machine learning models. In this example, Workgroup 1640 refers to the biotechnology area and Art Unit 1641 refers to the specific sub-field of immunology. The user is able to type in a word or phrase that needs to be expanded. The results of each model are displayed in a card with terms (words or phrases) that can either be up-voted (green triangle) or down-voted (red triangle). The up-voted terms are also added to a search string.

Every time a user up-votes or down-votes a term, it is remembered and the crowdsourcing models are updated. When a user initiates a new search, if the term was previously searched by one or more of her colleagues in the same Art Unit or Workgroup, those suggestions will be displayed in order based on votes. By allowing experts to fine tune, curate the correct terms, and by also manually add terms (for example, new cutting edge terms), there is a significant improvement in the quality of the results. Fig 4 shows the comparison of F1 scores before and after crowdsourcing for a sample set of words in the optics area. The average F1 score on the word embedding model for the of suggested relevant terms (F1 scores ranging from 0.5 to 0.25) did not approach the quality of those generated by optics area was 0.08391 vs. a score of 0.609 for the crowdsourcing model.

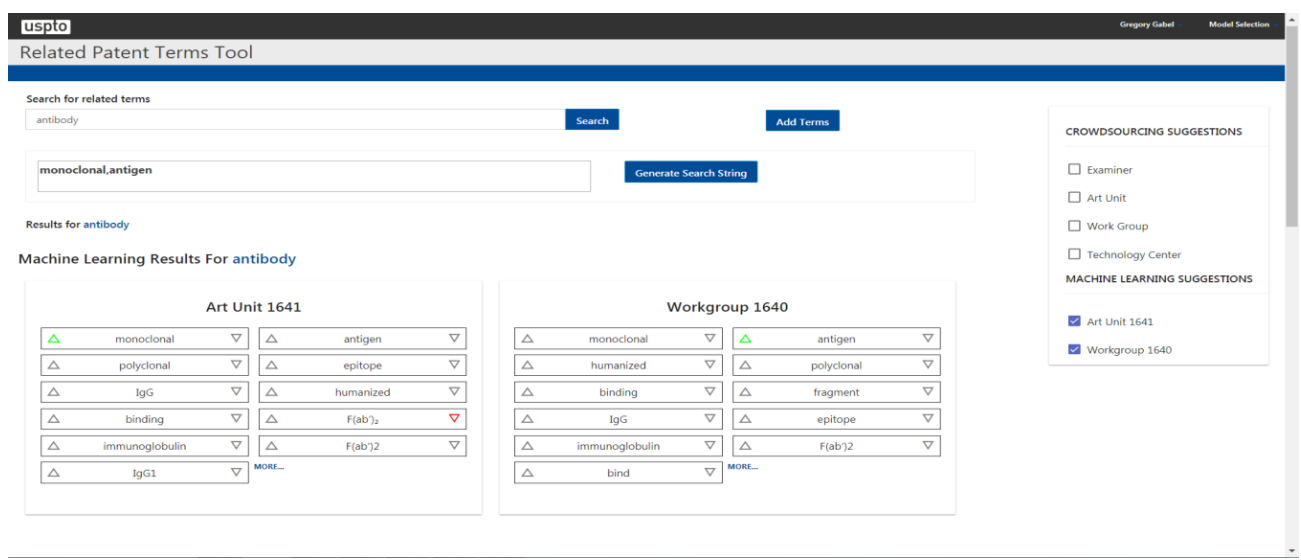

*Figure 3 User Interface that depicts how expert interactions can provide data for crowd sourcing models*

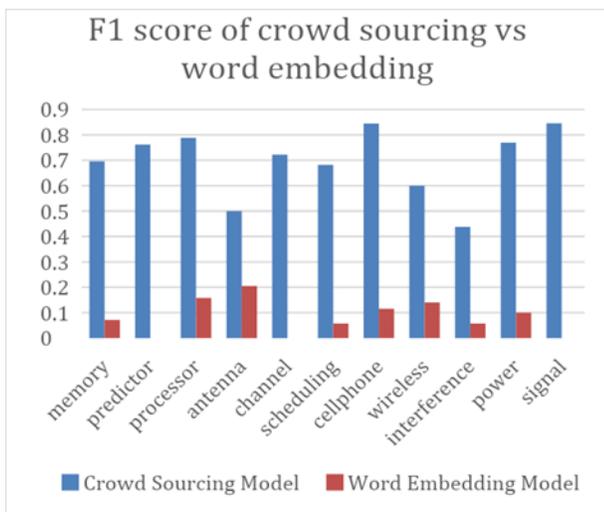

*Figure 4 F1 score of crowdsourcing results against PatSynSet compared with similar score for word embedding*

During this experiment, a small set of pilot users were able to interact with the system, with typically 3 to 4 users per area. However, the typical size of a crowd of examiners sharing that same technical field can range from 5-25 in specialized art units, to up to 500 examiners in high volume Art Units.

Our metrics for PatSynSet measured a narrow use case involving a limited number of users and entry terms. Deploying the crowdsourcing user interface to a larger examiner user group would allow us to greatly expand the data in PatSynSet. This combined approach of using machine learning results as seed data, and learning from the user interactions of professional users, is more tractable than either creating manually curated thesaurus (which is impractical to keep up to date) or only relying solely on automated word embedding based approaches.

## Lessons Learned

There is significant interest and potential in deploying Artificial Intelligence solutions to aid the mission of government institutions. Although AI shows enormous promise, the challenge of solving a complex problem such as patent retrieval at the caliber of an expert is daunting. State of the art solutions (Tannebaum et. al. 2015) have reported low scores on standardized IR tasks such as CLEF-IP and NTCIR (MAP scores of [ 0.05 , 0.15 ] in Prior Art Candidates search task). One of the first and most important lessons is to look for opportunities to recast the AI problem to make it more tractable. In this scenario, we have achieved this in several ways:

1. Identifying smaller-scoped problems, such as tackling just the query expansion aspect of patent searching, allows us to achieve results at a quality that can benefit experts right away.
2. AI is data-hungry, and yet the resources necessary to generate manually curated data may not be feasible for government agencies alone. We have overcome heavy upfront data costs by ensuring that our AI solutions are built around business processes that constantly involve user interactions which can be logged and learned from.
3. Accepting limits of current AI capabilities and building systems that allow humans to easily correct results suggested by AI. An AI solution does not necessarily need to be completely transparent, but it should provide enough information for an end user to be able to interact with and improve the results.

In the following section, we describe a solution that implements some of these lessons to creatively exploit user interactions for continuous improvement.

## Nearest Neighbors of Word Embedding Centroid

Retrieving synonyms for a patent term involves looking up nearest neighbors based on the word embedding model trained. This results in a static list of possible relevant terms. However, pursuing the spirit of allowing user interactions to guide the results, we can also re-compute related terms suggestions based on not just the initial patent term but also the results that user has shown a positive interest in. For example:

Results for term "lens"
```
lens | lenses, refracting, focal, aspherical,
aspheric, convex, biconvex, aspherically,
focusing, concave, doublet, cemented, zoom,
aberration, spherical, planoconcave,
biconcave, curvature, focus
```

Results for term "lens" when "optic" is up-voted by user
```
lens,optic | lenses, optical, fiber, lense,
refracting, optics, collimating, focusing,
aspheric, focal, aspherical, electro,
doublet, aspherically, spherical, convex,
assembly, converging
```

Results for term "lens" when "optic" and "microlens" is up-voted by user
```
lens, optic, microlens | lenses, optical,
collimating, GRIN, convex, lense, refracting,
microlensed, converging, lenslet, focal,
fiber, aspherically, concave, fresnel,
condensing
```

Results for term "lens" when "optic", "microlens" and "nanolens" is up-voted by user
```
lens, optic, microlens, nanolens | lenses,
microlenses, optical, collimating,
microlensed, microoptical, lense, GRIN,
fresnel, convex, refracting, converging,
lenslets, lenslet, fiber, diffractive
```

By mathematically computing the centroid of the word vectors chosen so far, and identifying the nearest neighbor of this centroid vector, we are able to obtain the most relevant suggestions. As examiners tend to choose highly related terms, the centroid based approach results in uncovering additional related terms. In the field of optics, the F1 scores of retrieved results improved as the number of chosen terms (lens - 1 vs. lens and optic - 2, lens, optic and microlens – 3, etc.) increased.

## Conclusion and Future Work

Tools to assist patent searchers with query expansion may increase recall of relevant prior art, improve consistency of the search process, and ultimately help assure the quality and validity of approved patents. In this work, we present a combined approach in which users are presented with initial suggestions of synonyms and other potentially relevant related terms for a given input term, as generated by word embedding models trained specifically on patent documents in the user's technological field. In most cases, use of a technology-specific corpus produced the best results. This approach is fast, automated, and may be readily scaled to additional technological fields.

Results were further augmented by implementing professional crowdsourcing, in which professional patent examiners proficient in query expansion assess and add to the machine learning results to create a database of expert-vetted results. Presently, users may interface with both word-embedding and crowdsourced models within a web-based user interface; and similar functionality may also be integrated within existing patent search platforms. Furthermore, as users make valid selections, based on the centroid of those selections, we are able to suggest more accurate related terms.

This semi-automated approach enables curation of a dynamic database of related terms useful in patent searching, while also facilitating knowledge sharing among patent examiners in related technological fields.

## Acknowledgements

Girish Showkatramani and Naresh Nula contributed code to make this work a reality. The authors would like to thank Matthew Such, Chris Chin, Elaine Greene, David Landrith, Aaron Pepe, Arva Adams, Rama Elluru, Tom Beach, Scott Williams, Jamie Holcombe and David Chiles for their guidance of this work. This work was made possible largely due to the support of Director Andrei Iancu and Deputy Director Laura Peter.

## References

Nanba, H. 2007. Query Expansion using an Automatically Constructed Thesaurus. *Proceedings of NTCIR-6 Workshop Meeting*, Tokyo, Japan.

Nanba, H., Kamaya, H., Takezawa, T., Okumura, M., Shinmori, A., Tanigawa, H. 2011. Automatic translation of scholarly terms into patent terms. *In Current Challenges in Patent Information Retrieval.* vol 29. Springer, Berlin, Heidelbergl, 373-388.

Konishi, K. 2005. Query Terms Extraction from Patent Document for Invalidity Search. *Proceedings of NTCIR-5 Workshop Meeting*, Tokyo, Japan.

Ganguly, D., Leveling, J., Magdy, W., and Jones, G. 2011. Patent query reduction using pseudo relevance feedback. *In Proceedings of the 20th ACM international conference on Information and knowledge management*, pp. 1953-1956.

Tannebaum W., Mahdabi P., Rauber A. 2015. Effect of Log-Based Query Term Expansion on Retrieval Effectiveness in Patent Searching. *CLEF* 2015. Lecture Notes in Computer Science, vol 9283. Springer, Cham.

Showkatramani, G., Krishna, A., Jin, Y., Pepe, A., Nula, N. and Gabel, G. 2018. User Interface for Managing and Refining Related Patent Terms. *In: Proceedings of HCI International 2018 – Posters*. 10.1007/978-3-319-92270-6_16.

Mikolov, T., Chen, K., Corrado, G., & Dean, J. 2013. *Efficient estimation of word representations in vector space.*

Bojanowski, P., Grave, E., Joulin, A., and Mikolov, T. 2017. Enriching Word Vectors with Subword Information. *In the Transactions of the Association for Computational Linguistics*.

Braslavski, P., Ustalov, D., Mukhin, M. 2014. A spinning wheel for YARN: user interface for a crowdsourced thesaurus. *In: Proceedings of the Demonstrations at the 14th Conference of the European Chapter of the Association for Computational Linguistics*, Gothenburg, Sweden, pp. 101–104.

Krek, S., Laskowski, C., Robnik-Šikonja, M. 2017. From Translation Equivalents to Synonyms: Creation of a Slovene Thesaurus Using Word co-occurrence Network Analysis. *In: Proceedings of eLex 2017*: Lexicography from Scratch, 19-21 September 2017, Leiden, Netherlands